\documentclass[a4paper,11pt]{article}

\usepackage{amssymb}
\usepackage{amsmath}
\usepackage{graphicx}
\usepackage{color}
\usepackage[a4paper]{geometry}
\usepackage{subfigure}
\usepackage[numbers]{natbib}
\usepackage{caption}
\usepackage{setspace}
\usepackage{epstopdf}
\usepackage{hyperref}
\usepackage{float}
\usepackage{booktabs}
\usepackage{multirow}
\usepackage{bm}

\hypersetup{
    pdftitle={Effect of different geometrically nonlinear strain measures on the static nonlinear response of isotropic and composite shells},
    pdfauthor={Rodolfo Azzara},
    pdfsubject={Effect of different geometrically nonlinear strain measures on the static nonlinear response of isotropic and composite shells},
    colorlinks=true,
    linkcolor=blue,
    citecolor=red,
    filecolor=magenta,
    urlcolor=blue
}

\newcommand{\vect}[1]{\boldsymbol{\mathbf{#1}}}

\title{Effect of different geometrically nonlinear strain measures on the static nonlinear response of isotropic and composite shells with constant curvature} 

\author{
A. Pagani$^{1}$\footnote{Associate professor. E-mail: alfonso.pagani@polito.it},
R. Azzara$^{1}$\footnote{PhD student. E-mail: rodolfo.azzara@polito.it},
B. Wu$^{2}$\footnote{Marie Curie Individual Fellow. E-mail: wubinlongchang@163.com},
E. Carrera$^{1}$\footnote{Professor of Aerospace Structures and
Aeroelasticity. E-mail: erasmo.carrera@polito.it}
\\ \\
$^{1}${\em Mul}$\,^2$ group \\
Department of Mechanical and Aerospace Engineering, Politecnico di Torino, \\
Corso Duca degli Abruzzi 24, 10129 Torino, Italy\\\\
$^{2}$School of Mathematics, Statistics and Applied Mathematics, \\ NUI Galway, University Road, Galway, Ireland
}

\date{}

\begin{document}
\maketitle

\noindent
{\bf Abstract:} 
{\em
The structural analysis of ultra-lightweight flexible shells and membranes may require the adoption of complex nonlinear strain-displacement relations. These may be approximated and simplified in some circumstances, e.g., in the case of moderately large displacements and rotations, in some others may be not. In this paper, the effectiveness of various geometrically nonlinear strain approximations such as the von Kármán strains is investigated by making use of refined shell formulations based on the Carrera Unified Formulation (CUF). Furthermore, geometrical nonlinear equations are written in a total Lagrangian framework and solved with an opportune Newton-Raphson method.
Test cases include the study of shells subjected to pinched loadings, combined flexure and compression, and post-buckling including snap-through problems. It is demonstrated that full geometrically nonlinear analysis accounting for full Green-Lagrange strains shall be performed whenever displacements are higher than the order of magnitude of the thickness and if compressive loads are applied.
}

\hrule
\section{Introduction}
Highly flexible structures are employed extensively in various engineering fields. The great potential of these structures is to exhibit large displacements/rotations without showing plastic deformations. This capability is an aspect of fundamental importance if linked to the industrial requirement to produce more advantageous structures in terms of cost and performance.

In this context, \textit{shell} structures have been widely used and studied over the years by researchers and scientists.
Shells consist of curved lightweight constructions, and they turn out to be very popular in structural engineering mainly owing to their characteristics of supporting external loads with high efficiency. Their outstanding mechanical properties are due to the curvature, which generates coupling between the membrane and the flexural behaviours, in both singly and doubly curved geometries. Furthermore, when sufficiently thin, shell structures can also undergo large displacements/rotations when extreme external loading conditions are applied, but they are still preserving post-buckling stiffness. 
The literature concerning theories of shells is vast \cite{novozhilov1959thin,calladine1989theory,niordson2012shell}.
The works of Poisson \cite{poisson1828memoire}, Love \cite{love1935mathematical}, Mindlin \cite{mindlin1951influence}, Kirchhoff \cite{kirchhoffdas}, Reissner \cite{reissner1945effect} and Cauchy \cite{cauchybook} represent the classical formulations available in the open literature.
These classical theories are typically adopted in commercial codes. Over the years, various higher-order two-dimensional (2D) theories were developed to overcome the drawbacks of the classical studies' hypotheses.  For instance, Reddy \cite{reddy1985higher} proposed a refined through-the-thickness kinematics, accounting for higher-order shear deformations, to study 2D composite structures. Mashat \textit{et al.} \cite{mashat2014evaluation} provided an assessment of the relevance of displacement variables in refined theories for isotropic and multi-layered shells, using an axiomatic/asymptotic technique. The refined theories of shell models were unified by Carrera in his early work, see \cite{Carre_AoCMiE_2003}. 
Readers are referred to other significant works on isotropic and composite shell theories \cite{li2019adaptable,Cinefra2016}, because the purpose of the article is not to provide a detailed description of such theories in the linear regime. 

In the context of large displacement/rotation fields, the geometrically nonlinear relations have to be considered to accurately predict the flexible structures' mechanical behaviour. This challenging problem is of fundamental importance for modern analysts to build an efficient and reliable numerical tool that is able to evaluate these nonlinearities in the mechanical response of shell structures.
Nonlinear analysis of isotropic and laminated shell structures is a topic of considerable research interest.
Since there is a large number of studies in the literature on this topic, readers are referred to the review provided by Budiansky \cite{budiansky1968notes}, Librescu and Schmidt \cite{librescu1988refined}, Palmiero \textit{et al.} \cite{palmieroref}. Many available works are focused on the theoretical developments and numerical solutions of shell models in the nonlinear regime. For example, Arciniega \textit{et al.} \cite{Arciniega2004} investigated bucking and post-buckling behaviours of shells under longitudinal compressions. 
Large deformation analysis of functionally graded (FG) shells were presented by the same author \cite{Arciniega2007} adopting tensor-based finite element formulations for geometrically nonlinear relations.
Palazotto \cite{palazotto1992nonlinear} provided a detailed description of geometrically nonlinear theories for shell models with transverse shear deformations, emphasizing the nonlinear composite shell behaviour.
Nonlinear analysis of shells based on the degenerated isoparametric shell element was studied by Kuo-Mo and Yeh-Ren \cite{kuo1989nonlinear}. Hajlaouia \textit{et al.} \cite{hajlaoui2017nonlinear} formulated a higher-order shear strain enhanced solid-shell element to perform an accurate nonlinear dynamic analysis of 2D FG materials and structures.
Sze \textit{et al.} \cite{Sze2004} provided significant benchmark solutions for the analysis of various geometrically nonlinear shell structures in a recent work.

Most of the available researches employ the classical nonlinear von Kármán strains to perform geometrically nonlinear analysis of highly flexible structures. 
For example, Ma and Wang \cite{Ma2003} adopted the von Kármán nonlinearity to study large deflections of 2D FG circular structures under mechanical and thermal loadings. Zao and Liew \cite{zhao2009geometrically} considered nonlinear formulations based on von Kármán strains to conduct the nonlinear response analysis of FG ceramic-metal shells. Arumugam and Reddy \cite{arumugam2018nonlinear} adopted the von Kármán strains to carry out nonlinear analysis of ionic polymer-metal laminated elements.
Geometrically nonlinear analyses of piezo-laminated smart shells were presented by Kulkarni and Bajoria \cite{kulkarni2007large} by employing higher-order shear deformation formulations and adopting von Kármán hypotheses.
Furthermore, their commercial codes adopted these classical von Kármán nonlinear strain theories.
As mentioned by Carrera and Parisch \cite{Carrera1997}, the classical von Kármán theory showed reliable results
in the deflection analysis of thin shells when displacements are the same order of magnitude of the thickness.
In contrast, less accuracy was obtained in the case of thick structures, as confirmed by the same authors. Moreover, the inaccuracy predicted by the von Kármán theory is accentuated when shear loadings are considered, in which case deformations exhibit large rotations. In addition, the displacements are overestimated when using the von Kármán's nonlinear theories in the advanced nonlinear regime, as reported by Kim and Chaudhuri \cite{Kim1995}. 
In the advanced nonlinear domain, the proposed method in this paper aims to compare different geometrically nonlinear strain approximations for large-deflection and post-buckling nonlinear response analyses of isotropic or composite shells. The goal of the present research is to provide benchmark solutions on shell structure problems, pointing out where the classical von Kármán nonlinear strain approximation is accurate and where it is not. In other words, we will present different prediction results obtained when different geometrically nonlinear strains are adopted in moderate and large displacement/rotation fields.

The Carrera Unified Formulation (CUF) was implemented to perform geometrically nonlinear analyses of isotropic and laminated plates and shells in the last years \cite{wu2019geometrically,carrera2020popular,pagani2019evaluation,carrera2020vibration, pagani2020accurate,pagani2021stress,wu2019large,pagani2020evaluation} and characterizes the foundations of the proposed method. 
In the framework of CUF, any degree of refinement of a model can be obtained since the adopted theory's order is treated as an analysis input. 
According to the textbooks \cite{CarreGP_2011,carrera2014finite}, any theory is degenerated into generalized kinematics employing arbitrary expansions. In this article, in particular, Lagrange expansion (LE) are considered, which may eventually lead to a layerwise (LW) approach for the analysis of laminated structures, see \cite{Carre_AJ_1999,carrera1999multilayered2}. Basically, the nonlinear governing equations and the
related finite element (FE) array of any model are formulated using the \textit{fundamental nuclei} (FN). The FNs represent the basic building blocks of the presented formulations.
Furthermore, different geometrically nonlinear refined shell theories from full Green-Lagrange (GL) strains to the classical von Kármán strains are automatically and opportunely obtained by adopting the CUF due to its intrinsic scalable nature.
Once the nonlinear governing equations are derived, the path-following Newton-Raphson linearization method based on the arc-length constraint is employed to compute the geometrically nonlinear solutions. 

This paper is structured as follows: (i) first, Section \ref{F} discusses some information about the related geometrically nonlinear relations and the 2D CUF shell model; (ii) next, the numerical results are given in Section \ref{Z}, where particular interest focuses on the comparison between different results obtained by adopting the full Green-Lagrange nonlinear strain tensor and the classical von Kármán nonlinear theories; (iii) finally, Section \ref{V} makes the conclusions.
\section{Geometrically nonlinear refined shell models} \label{F}
\subsection{Preliminaries}
\begin{figure}[htbp]
\centering
\includegraphics[scale=0.65, angle=0]{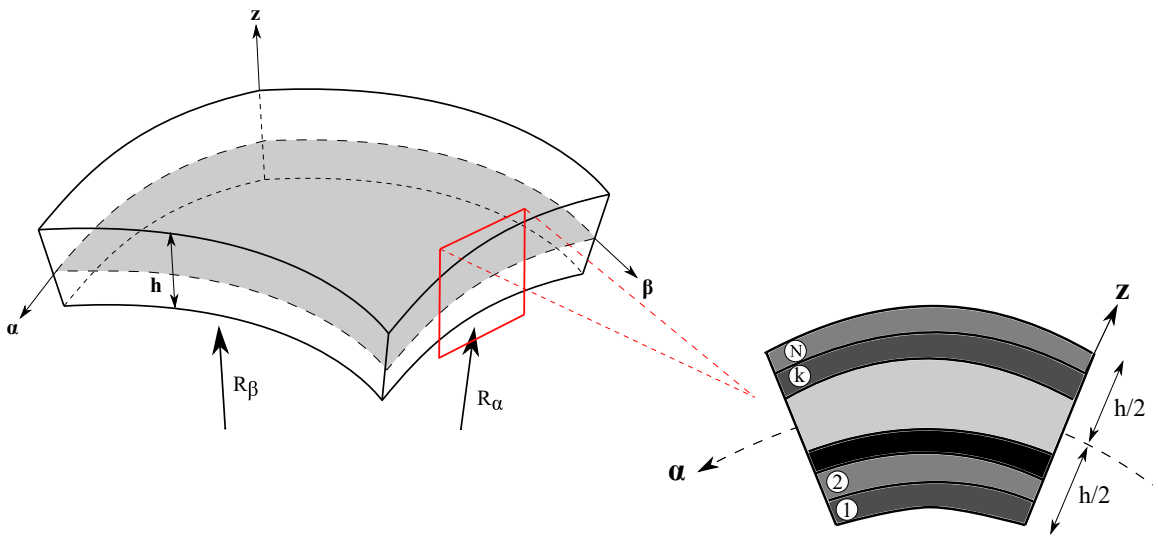}
\caption{Generic laminated composite shell model and related curvilinear coordinate system.}\label{fig 305}
\end{figure}

The discussion is presented by taking into account a $N$-ply laminated composite shell structure, as illustrated in Fig. \ref{fig 305}, with the in-plane $\alpha$-$\beta$ and through-the-thickness $z$ domains, in which $k$ indicates the $k^{th}$ layer. For the sake of clarity, the same formulations are applied to homogeneous isotropic models.
The typical three-dimensional displacement vector of a point in the composite shell is:
\begin{equation}
\bm{u}^{k}(\alpha,\beta,z)=\{\;u_{\alpha}^{k}\;u_{\beta}^{k}\;u_{z}^{k}\;\}^{\text{T}}
\end{equation}
In addition, the Green-Lagrange (GL) strain, $\bm{\epsilon}$, and the second Piola-Kirchhoff (PK2) stress, $\bm{\sigma}$, vectors for each layer $k$ are written as:
\begin{equation}
\begin{array}{l}
\bm{\epsilon}^{k}=\{\epsilon_{{\alpha}{\alpha}}^{k},\;\epsilon_{{\beta}{\beta}}^{k},\;\epsilon_{zz}^{k},\;\epsilon_{{\alpha}z}^{k},\;\epsilon_{{\beta}z}^{k},\;\epsilon_{{\alpha}{\beta}}^{k}\}^{\text{T}}\\\\
\bm{\sigma}^{k}=\{\sigma_{{\alpha}{\alpha}}^{k},\;\sigma_{{\beta}{\beta}}^{k},\;\sigma_{zz}^{k},\;\sigma_{{\alpha}z}^{k},\;\sigma_{{\beta}z}^{k},\;\sigma_{{\alpha}{\beta}}^{k}\}^{\text{T}}
\end{array}
\end{equation}
In the regime of large displacement and rotation fields, the problem of defining the strain tensor from the initial undeformed state is not easy \cite{Malvern1969}. For this reason, accurate definitions of strain and stress tensors are necessary to perform reliable nonlinear analysis.
The Lagrangian formulations are typically adopted in the pure geometrically nonlinear analysis due to its natural undeformed state, to which the structure returns when unloaded. By using a Lagrangian approach, strains are formulated in terms of the undeformed configuration. On the contrary, strains are expressed as functions of the deformed configuration in the Eulerian description. A large number of advantages are entailed when a Lagrangian method is adopted \cite{Pai2007book}.

The proposed approach is implemented using the \textit{total Lagrangian} formulations and employs the GL strains. In essence, the GL strain vector is written as:
\begin{equation}
\bm{\epsilon}^{k}= \bm{\epsilon}_{l}^{k}+\bm{\epsilon}_{nl}^{k}= (\bm{b}_{l}^{k}+\bm{b}_{nl}^{k})\bm{u}^{k}
\end{equation}
in which $\bm{b}_{l}$ and $\bm{b}_{nl}$ represent the 6$\times$3 linear and nonlinear differential operators. 
Complete forms of these two matrices for the 2D shell model are given in the following \cite{wu2019geometrically}:
\begin{equation} \label{eq}
\resizebox{\linewidth}{!}{%
$
\vect{b}_l = \left[\begin{array}{ccc}
\dfrac{\partial_{\alpha}} {H_{\alpha}} & 0 & \dfrac{1} {H_{\alpha}R_{\alpha}} \\\\
0 & \dfrac{\partial_{\beta}} {H_{\beta}} & \dfrac{1} {H_{\beta}R_{\beta}} \\\\
0 & 0 & \partial_z \\\\
\partial_z - \dfrac{1} {H_{\alpha}R_{\alpha}} & 0 & \dfrac{\partial_{\alpha}} {H_{\alpha}} \\\\
0 & \partial_z - \dfrac{1} {H_{\beta}R_{\beta}} & \dfrac{\partial_{\beta}} {H_{\beta}} \\\\
\dfrac{\partial_{\beta}} {H_{\beta}} & \dfrac{\partial_{\alpha}} {H_{\alpha}} & 0
\end {array}\right]\text{,} \qquad
\vect{b}_{nl} = \left[\begin{array}{ccc}
P_{11}\dfrac{1}{2H_{\alpha}^2}\left[ \left(\partial_{\alpha}\right)^2 + \dfrac{2u_z{\partial_{\alpha}}}{R_{\alpha}} + \dfrac{u_{\alpha}}{R_{\alpha}^2}\right] & P_{12}\dfrac{\left(\partial_{\alpha}\right)^2}{2H_{\alpha}^2} & P_{13}\dfrac{1}{2H_{\alpha}^2}\left[ \left(\partial_{\alpha}\right)^2 - \dfrac{2u_{\alpha}{\partial_{\alpha}}}{R_{\alpha}} + \dfrac{u_{z}}{R_{\alpha}^2}\right]\\\\
P_{21}\dfrac{\left(\partial_{\beta}\right)^2}{2H_{\beta}^2} & P_{22}\dfrac{1}{2H_{\beta}^2}\left[ \left(\partial_{\beta}\right)^2 + \dfrac{2u_z{\partial_{\beta}}}{R_{\beta}} + \dfrac{u_{\beta}}{R_{\beta}^2}\right] & P_{23}\dfrac{1}{2H_{\beta}^2}\left[ \left(\partial_{\beta}\right)^2 - \dfrac{2u_{\beta}{\partial_{\beta}}}{R_{\beta}} + \dfrac{u_{z}}{R_{\beta}^2}\right] \\\\
P_{31}\dfrac{1}{2} \left(\partial_z\right)^2 & P_{32}\dfrac{1}{2} \left(\partial_z\right)^2 & P_{33}\dfrac{1}{2} \left(\partial_z\right)^2 \\\\
P_{41}\dfrac{1}{H_{\alpha}}\left(\partial_ {\alpha} \, \partial_z + \dfrac{u_z{\partial_{z}}}{R_{\alpha}}\right) & P_{42} \dfrac{\partial_{\alpha} \, \partial_z}{H_{\alpha}} & P_{43}\dfrac{1}{H_{\alpha}}\left(\partial_ {\alpha} \, \partial_z - \dfrac{u_{\alpha}{\partial_{z}}}{R_{\alpha}}\right) \\\\
P_{51}\dfrac{\partial_{\beta} \, \partial_z}{H_{\beta}} & P_{52}\dfrac{1}{H_{\beta}}\left(\partial_ {\beta} \, \partial_z + \dfrac{u_z{\partial_{z}}}{R_{\beta}}\right) & P_{53}\dfrac{1}{H_{\beta}}\left(\partial_ {\beta} \, \partial_z - \dfrac{u_{\beta}{\partial_{z}}}{R_{\beta}}\right) \\\\
P_{61}\dfrac{1}{H_{\alpha}H_{\beta}}\left(\partial_{\alpha} \, \partial_{\beta} + \dfrac{u_z{\partial_{\beta}}}{R_{\alpha}} + \dfrac{u_{\beta}}{R_{\alpha}R_{\beta}}\right) & P_{62}\dfrac{1}{H_{\alpha}H_{\beta}}\left(\partial_{\alpha} \, \partial_{\beta} + \dfrac{u_z{\partial_{\alpha}}}{R_{\beta}}\right) & P_{63}\dfrac{1}{H_{\alpha}H_{\beta}}\left(\partial_{\alpha} \, \partial_{\beta} - \dfrac{u_{\alpha}{\partial_{\beta}}}{R_{\alpha}} - \dfrac{u_{\beta}{\partial_{\alpha}}}{R_{\beta}}\right)
\end {array}\right]%
$
}
\end{equation} 
where $\partial_{\alpha}=\partial(\cdot)/\partial\alpha$, $\partial_{\beta}=\partial(\cdot)/\partial\beta$, $\partial_{z}=\partial(\cdot)/\partial{z}$, $P_{ij}(i=1-6, j=1-3)$ represent  the parameters used to opportunely simplify or tune the nonlinear strain measures and $H_{\alpha}$, $H_{\beta}$ are defined as:
\begin{equation}
H_{\alpha}^{k}= A^{k}(1+z_{k}/R_{\alpha}^{k}), \;\;\;\;\;\; H_{\beta}^{k}= B^{k}(1+z_{k}/R_{\beta}^{k}), 
\end{equation}
where $R_\alpha^{k}$ and $R_\beta^{k}$ indicate the radii of the middle surface of the layer $k$ and $A^{k}$ and $B^{k}$ denote the Lamé parameters of the in-plane mid-surface. This paper only considers the shells with constant curvatures and the in-plane coordinates $\alpha$ and $\beta$ are chosen to be the arc-length coordinates, therefore $A^{k}$= $B^{k}$= 1.
A complete review of geometrically nonlinear shell formulations is not the purpose of this work. Further details can be found in \cite{reddy2003mechanics,green1992theoretical}. 

\begin{figure}[htbp]
\centering
\includegraphics[scale=0.85, angle=0]{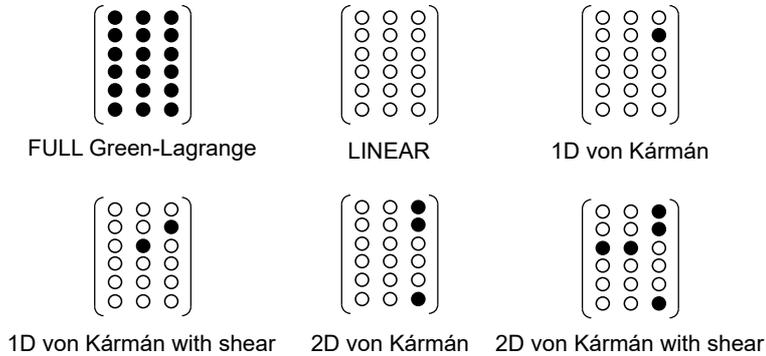}
\caption{Different geometrically nonlinear strain models for shells.}\label{fig 7620}
\end{figure}

Researchers have developed many approximate geometrically nonlinear models for 2D and one-dimensional (1D) structures from simplifications of the three-dimensional (3D) full geometrical relations over the years. The von Kármán strain theory for 2D shells, see \cite{GolDenveizer1961}, represents a classical and well-known example.
In the domain of moderate rotations, the hypotheses of 2D von Kármán models state that only the nonlinear terms of Eq. \ref{eq} associated with the in-plane partial derivatives of the transverse displacement cannot be neglected. 
Thus, the only non-zero components of geometrically nonlinear strains are $P_{13}$, $P_{23}$, and $P_{63}$ $\not =$ 0.
If, adding the nonlinear shear effects as well, the additional non-zero terms that should be considered are $P_{31}$ and $P_{32}$.
%
%
Instead, according to the von Kármán hypotheses, in the case of 1D models, the only non-zero component is $P_{23}$ $\not =$ 0. In the case of von Kármán strain theory for 1D models with nonlinear shear effects, the $P_{32}$ strain component should be also considered.
In summary, different geometrically nonlinear strain models compared in this paper are shown in Fig. \ref{fig 7620}, where the black dots indicate the activated nonlinear strain components with reference to the matrix $\vect{b}_{nl}$ in Eq \ref{eq}.
%
%

The present work aims to investigate the effect of various geometrically nonlinear strain models on the static nonlinear response of metallic and laminated shells in moderate to large displacement and rotation fields, emphasizing the comparison between 3D full geometrically nonlinear strain relations and 2D simplified von Kármán models.

For linear elastic materials, the constitutive relations are expressed as:
\begin{equation}
\bm{\sigma}^{k}= \bm{\tilde{C}}^{k}\bm{\epsilon}^{k}
\end{equation}
in which the complete expressions of the material elastic matrix $\bm{\tilde{C}}$ are provided in \cite{Bathe_1996,hughes2012finite}.

In the framework of Carrera Unified Formulation, the nonlinear governing equations of isotropic and laminated composite shells are formulated in a unified manner due to the scalable nature of the CUF.
Furthermore, it is possible to formulate various geometrically nonlinear models by cancelling or adding different nonlinear strain terms from or into the CUF fundamental nuclei (FN).
\subsection{Carrera Unified Formulation (CUF)}
In the framework of the 2D shell CUF, the 3D displacement field is defined as:
\begin{equation}
\bm{u}^{k}({\alpha},{\beta},z)= F_{\tau}^{k}(z)\bm{u}_{\tau}^{k}({\alpha},{\beta}) \;\;\;\;\;\;\; \tau= 0,1,...,N
\end{equation}
in which $\bm{u}_{\tau}({\alpha},{\beta})$ represents the generalized displacement vector depending on the in-plane coordinates $\alpha$ and $\beta$, $k$ denotes the layer index, $N$ indicates the order of expansion in the thickness direction and $F_{\tau}$ are the expansion functions of the thickness coordinate $z$.
It is possible to choose $F_{\tau}$ and $N$ arbitrarily, which defines the class of the 2D CUF shell model.
For a detailed mathematical derivation of the shell FE formulations in the CUF framework, readers are referred to \cite{carrera2014finite}.

In this article, the LW \cite{Carre_AJ_1999,carrera1999multilayered2,reddy1984simple,reddy1987generalization} approach based on the LE model is used. 
LW theories divide and expand the displacement field within each material layer. The continuity of the displacements is guaranteed at the interface level by the continuous expansion functions, to have an accurate evaluation of the deformation and stress distributions. By doing so, the homogenization is carried out at the interface layer. 
When LE models are considered, the unknowns are only displacements; i.e., they have only displacements as degrees of freedom. The displacements at each interface obey the compatibility conditions.
Readers are referred to \cite{CarrePP_AJ_2013,CarreP_JoSaV_2014,CarreP_AJ_2016} for more details about the Lagrange polynomials along the thickness direction used in this paper.
The acronym LDN denotes the adopted CUF shell formulations, which represent the LE of order $N$. Essentially, the two-node linear (LD1), three-node
quadratic (LD2), and four-node cubic (LD3) Lagrange expansion functions are employed in the
thickness direction to obtain linear to higher-order kinematics CUF shell theories.

Independent of the shell model kinematics, the finite element method (FEM) is used to discretize the generalized displacement vector in the $\alpha$-$\beta$ plane, as follows:
\begin{equation}
\bm{u}_{\tau}^{k}({\alpha},{\beta})=N_{i}(\alpha,\beta)\bm{q}_{{\tau}i}^{k} \;\;\;\;\;\;\; i=1,2,...,n_{el}
\end{equation}
in which $N_{i}(\alpha,\beta)$ are the shape functions, $\bm{q}_{{\tau}i}$ represents the unknown nodal parameters, $n_{el}$ is the number of nodes per element and the repeated index \textit{i} indicates summation.
For clarity, the classical 2D nine-node quadratic (Q9) FE will be considered as the shape function in the following analysis. During the analyses, no problems related to membrane and shear locking (MITC9) were found. For the sake of completeness, readers are referred to \cite{Cinefra2016,CinefVC_IJoSaNM_2015,carrera2016mitc9,cinefra2013mitc9} for a detailed description of the strategy to contrast the membrane and shear locking phenomenon for a nine-node shell element in the CUF domain.

The principle of virtual work is utilized to derive the formulations of the nonlinear FE governing equations. In essence:
\begin{equation}
{\delta}L_{int}={\delta}L_{ext}
\end{equation}
where ${\delta}L_{int}$ and ${\delta}L_{ext}$ represent the virtual variations of the strain energy and the work of external loadings. They are expressed as follows:
\begin{equation}
\begin{array}{l}
{\delta}L_{int}= {\delta}\bm{q}_{sj}^T\bm{K}_{0}^{ij{\tau}s}\bm{q}_{{\tau }i} + {\delta}\bm{q}_{sj}^T\bm{K}_{lnl}^{ij{\tau}s}\bm{q}_{{\tau}i} + {\delta}\bm{q}_{sj}^T\bm{K}_{nll}^{ij{\tau}s}\bm{q}_{{\tau}i} + {\delta}\bm{q}_{sj}^T\bm{K}_{nlnl}^{ij{\tau}s}\bm{q}_{{\tau}i} \\\\
\;\;\;\;\;\;\;\;\; = {\delta}\bm{q}_{sj}^T\bm{K}_{S}^{ij{\tau}s}\bm{q}_{{\tau}i}
\end{array}
\end{equation}
in which $\bm{K}_{S}^{ij{\tau}s}$ represents the secant stiffness matrix, $\bm{K}_{0}^{ij{\tau}s}$ is the linear contribution and $\bm{K}_{lnl}^{ij{\tau}s}$, $\bm{K}_{nll}^{ij{\tau}s}$ and $\bm{K}_{nlnl}^{ij{\tau}s}$ indicate the nonlinear contribution. 
These components are written in the form of FNs. 
Instead, the virtual variation of the external work can be formulated as:
\begin{equation}
{\delta}L_{ext}= {\delta}\bm{q}_{sj}^{T}\bm{p}_{sj}
\end{equation}

After some mathematical operations, the nonlinear equilibrium equations read as:
\begin{equation} 
\bm{K}_{S}^{ij{\tau}s}\bm{q}_{{\tau}i}-\bm{p}_{sj}=0
\label{eqnr1}
\end{equation} 
which is a set of three nonlinear algebraic equations, where $\bm{K}_{S}^{ij{\tau}s}$ indicates the \textit{secant} stiffness matrix and $\bm{p}_{{\tau}i}$ represents the nodal loading vector. Note that $\bm{K}_{S}^{ij{\tau}s}$ is expressed by means of FN, which is a $3 \times 3$ matrix that can be expanded by looping the indexes $i$, $j$, ${\tau}$ and $s$.
The complete forms of $\bm{K}_{S}^{ij{\tau}s}$ and $\bm{p}_{{\tau}i}$ are omitted here, see \cite{wu2019geometrically,pagani2019evaluation,Pagani2018,wu2019accurate} for a detailed description. Equation \ref{eqnr1} is arbitrarily expanded to obtain any desired theory by choosing the value for $\tau,s= 1,2,...,N$ and $i,j=1,2...,n_{el}$ to give:
\begin{equation}
\bm{K}_{S}\bm{q}-\bm{p}=0
\label{eqnr2}
\end{equation} 
where $\bm{K}_{S}$, $\bm{q}$ and $\bm{p}$ represent the global, assembled FE arrays of the final structure.

Finally, the path-following Newton-Raphson linearization method (or tangent method) \cite{Carre_Cs_1994,Crisf_CS_1981} is chosen to solve the nonlinear system. According to the Newton-Raphson method, Eq. \ref{eqnr2} is expressed as follows:
\begin{equation}
\bm{\varphi}_{res}= \bm{K}_{S} \bm{q} - \bm{p}=0
\label{EQ3}
\end{equation} 
where $\bm{\varphi}_{res}$ denotes the vector of the residual nodal forces (unbalanced nodal force vector).
Equation \ref{EQ3} can be linearized by expanding $\bm{\varphi}_{res}$ in Taylor's series about a known solution ($\bm{q},\bm{p}$). By introducing the tangent stiffness matrix $\bm{K}_{T}$ and assuming that the load varies directly with the vector of the reference loadings $\bm{p}_{ref}$, that it has a rate of change equal to the load parameter $\lambda$, i.e., $\bm{p}=\lambda\bm{p}_{ref}$, we obtain in compact form:
\begin{equation}
\bm{K}_{T}\delta\bm{q}=\delta\lambda\bm{p}_{ref}-\bm{\varphi}_{res}
\end{equation} 
Since the load-scaling parameter $\lambda$ is taken as a variable, an additional governing equation is required and this is given by a constraint relationship to finally obtain:
\begin{equation} 
\begin{cases}
\bm{K}_{T}\delta\bm{q}=\delta\lambda\bm{p}_{ref}-\bm{\varphi}_{res}\\
c(\delta\bm{q},\delta\lambda)=0
\end{cases}
\end{equation}
For the sake of brevity, readers are referred to \cite{Pagani2018,PaganC_CS_2017} for the complete description.

\section{Numerical results} \label{Z}
In this paper, representative benchmark problems accounting for large displacements and rotations are discussed. Particular emphasis focuses on the effectiveness of various geometrically nonlinear strain-displacement relations for different structure and loading problems.
Both isotropic and laminated composite shell structures are investigated.
It is important to underline that the curvature is one of the main parameters to consider in the analysis of shell structures. However,  only shells with constant curvature are analyzed in this research. The effect of the curvature and doubly-curved shell structures will be investigated in future works. 
In this regard, convergence analyses are conducted first.
Then, based on the converged model, large-deflection equilibrium curves of various 2D CUF shell models are presented. A comparison between the 3D full Green-Lagrange nonlinear strains and the classical von Kármán strains is reported. For completeness, readers are referred to \cite{carrera2021buckling} for a comparison between the linear buckling results and those obtained by incremental solutions. 
\subsection{Isotropic pinched thin-walled cylindrical shell}
\begin{figure}[htbp]
\centering
\includegraphics[scale=0.65, angle=0]{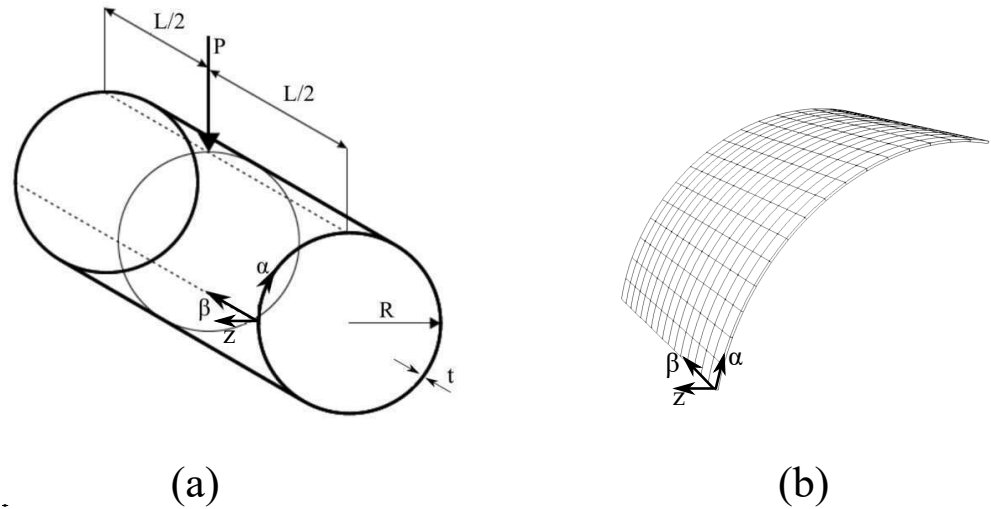}
\caption{Representation of a) the entire pinched thin-walled cylindrical shell under a transverse load and b) the in-plane mesh approximation of the model studied.}\label{fig 001}
\end{figure}

A clamped-clamped pinched thin-walled cylindrical shell under a transverse load is considered as the first example, see Fig. \ref{fig 001}. Owing to the symmetry of the structure, only one-eighth of the entire model is analyzed.
The geometrical and material data come from the book of Fl{\"u}gge \cite{flugge2013stresses}. The investigated model has the following characteristics: $L$ = 600 in, $t$ = 3 in and $R$ = 300 in. The material data are: $E$ = 3$\times$10$^6$ psi, $\nu$ = 0.3. 
The structure is subjected to large deflections due to a transverse load $P$ applied as in Fig. \ref{fig 001}.

\begin{figure}[htbp]
\centering
\subfigure [Mesh approximation]
{\includegraphics[scale=0.35, angle=0]{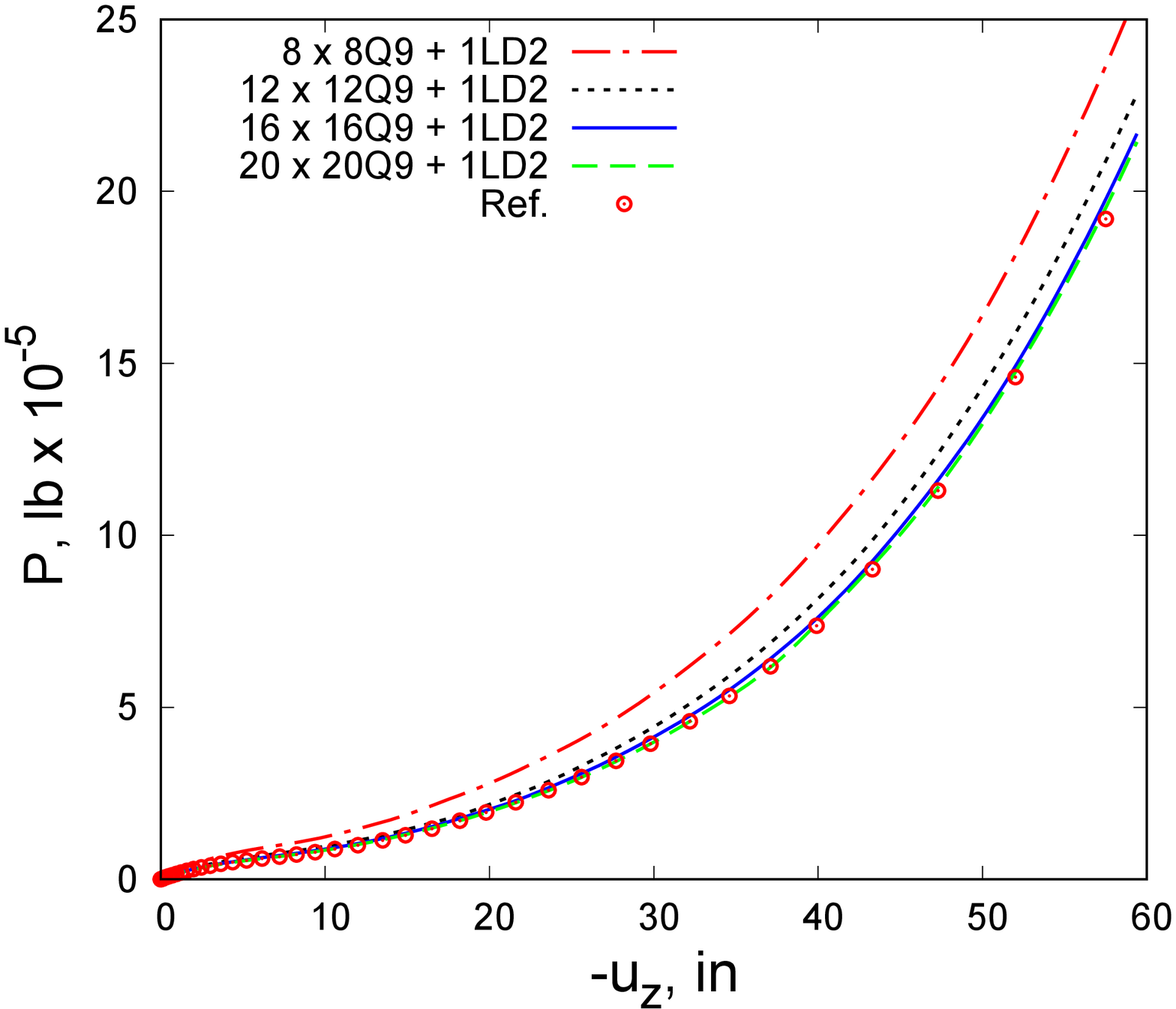}}
\subfigure [Kinematic expansion]
{\includegraphics[scale=0.35, angle=0]{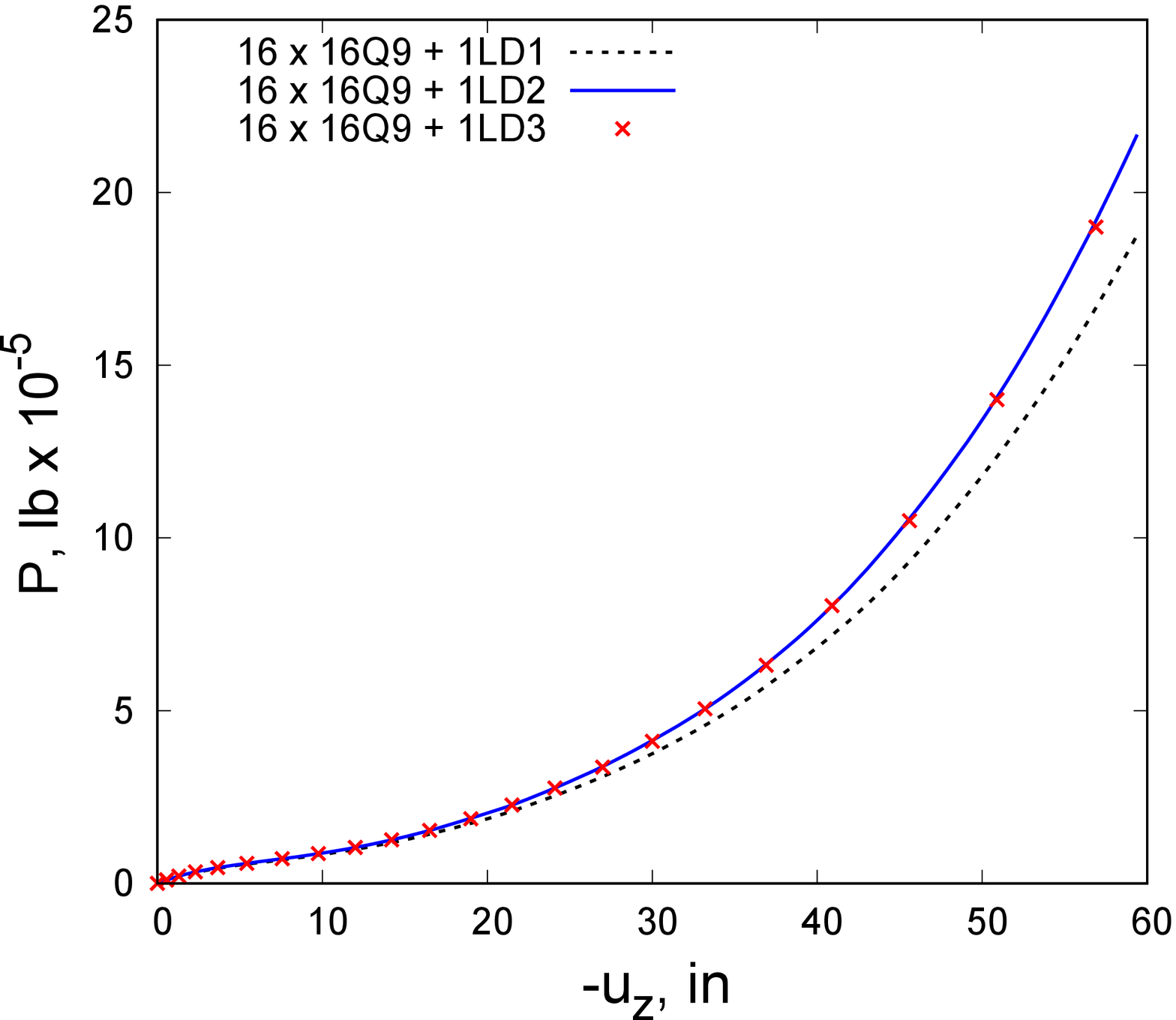}}
\caption{Convergence study of the nonlinear response curves for the pinched thin-walled cylindrical shell on both ends. Comparison of (a) various in-plane mesh approximations and (b) different orders of Lagrange expansion functions in the thickness direction.}\label{fig 050}
\end{figure}

First of all, convergence analyses on the in-plane finite element mesh and different kinematic expansion functions in the thickness direction are needed to perform an accurate comparison between the numerical results of different nonlinear strain approximations.
Specially, we evaluate the nonlinear equilibrium path of the clamped-clamped pinched thin-walled cylindrical shell for different in-plane meshes and LE functions in the thickness direction.
Figure \ref{fig 050}a compares the transverse deflection at the loading point for different 2D CUF shell elements, where the in-plane meshes from 64Q9 to 400Q9 FEs are employed, while the evaluations adopting one LD1, LD2 or LD3 are illustrated in Fig. \ref{fig 050}b. 
The proposed method was validated by comparing the results with those provided by Pagani \textit{et al.} \cite{pagani2019evaluation}, in which the same structure was studied based on the higher-order 1D beam elements.
Moreover, the transverse displacement values for various CUF shell models and loads, along with the total degrees of freedom (DOFs), are tabulated in Table \ref{Table050}.
\begin{table}[htbp]
\centering
\begin{tabular}{cccccc}
\toprule
\toprule 
\multirow{3}{*}{ CUF shell model}  & \multirow{3}{*}{DOFs}  &   \multicolumn{2}{c}{-$u_{z}$ [in]} \\
\cmidrule(l){3-4} 
 &  & 2.5$\times$10$^5$ lb & 15$\times$10$^5$ lb  \\
\midrule
8  x 8Q9  + 1LD2  & 2601  & 18.5 & 48.1  \\
12 x 12Q9 + 1LD2  & 5625  & 21.8 & 50.8  \\
16 x 16Q9 + 1LD2  & 9801  & 22.9 & 52.0   \\
20 x 20Q9 + 1LD2  & 15129 & 22.9 & 52.1    \\
Ref. \cite{pagani2019evaluation}  & 10920 & 23.1 & 52.4  \\
\bottomrule
\bottomrule
\end{tabular} 
\caption{Equilibrium points of nonlinear response curves of the pinched cylindrical shell for various models and loads at the loading point.}\label{Table050}
\end{table}
As evident from Fig. \ref{fig 050} and Table \ref{Table050}, to carry out an accurate static nonlinear response analysis and demonstrate the effects of the various nonlinear strain terms below, the structure should be modelled by employing 16$\times$16Q9 for the in-plane mesh approximation and only one LD2 in the thickness direction.

\begin{figure}[htbp]
\centering
\includegraphics[scale=0.55, angle=0]{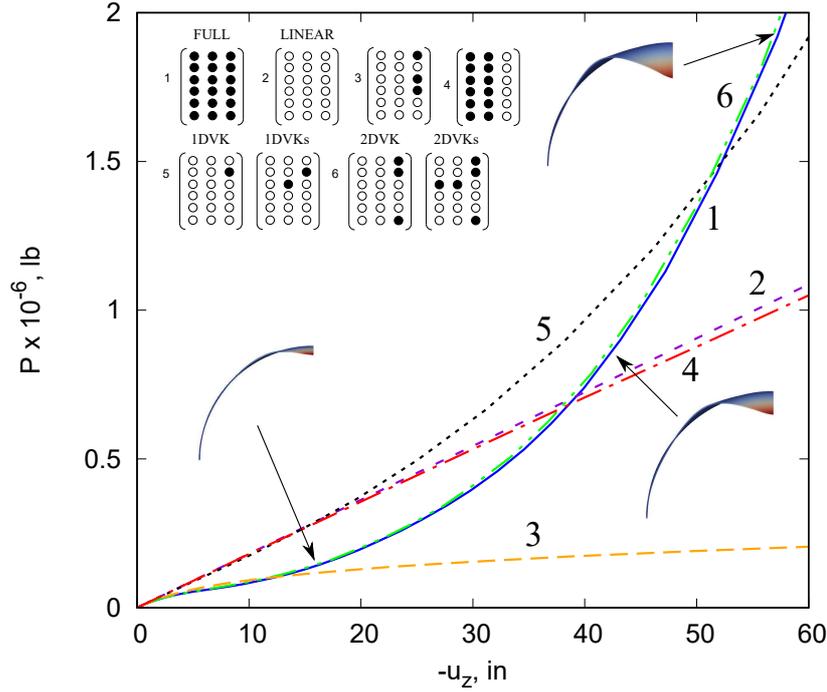}
\caption{Equilibrium curves evaluated at loading point of the pinched thin-walled cylindrical shell under a transverse load for different geometrically nonlinear approximations.}\label{fig 002}
\end{figure}

Figure \ref{fig 002} depicts the equilibrium curves of the pinched cylindrical shell under a transverse load for various geometrically nonlinear strain models at the loading point. In addition, some deformed configurations are illustrated in the same figure.
Basically, different nonlinear strain terms of the operator $\bm{b}_{nl}$ in Eq. \ref{eq} are activated in each geometrically nonlinear CUF shell theory.
In essence, black dots in Fig. \ref{fig 002} denote the activated nonlinear strain terms. 
For example, the analysis with all nonlinear terms involved (i.e., 3D full GL nonlinear strain) is abbreviated as the ``Full", whereas that with all nonlinear terms excluded indicates ``linear" analysis. The ``1DVK" and ``2DVK" analyses represent the von Kármán assumptions corresponding to 1D and 2D models. Instead, the analyses including nonlinear shear effects will be referred to as ``1DVKs" and ``2DVKs".
Analysis case number 4 points out that neglecting the higher-order derivatives of the transverse displacement component u$_{z}$ (the third column of matrix $\bm{b}_{nl}$) affects the accuracy of the nonlinear solution. In fact, the variation trend is very similar to that of linear analysis (case number 2).
As for the von Kármán models, case numbers 5 and 6 are the solutions for the traditional 1D and 2D sets of approximations. In particular, the 2DVK curve is very similar to the 3D full geometrically nonlinear solution also for moderate and large displacements/rotations. In contrast, the 1DVK case leads to a remarkably different curve than the FULL theory. Especially, this curve follows the linear solution until the transverse displacement reaches the value of $20$ in. After that, the displacement predicted by the 1DVK (and 1DVKs) is larger than the linear case. It is important to underline that including or not the nonlinear shear terms in the von Kármán approximation leads to nearly the same results.
For the sake of completeness of readers, adopting a symmetrical model, for example, one-eighth of the entire cylinder as in this case, may not provide accurate results when considering non-symmetrical load conditions or composite structures with complex laminations.
\subsection{Composite cylindrical shell subjected to compressive and transverse loads}
\begin{figure}[htbp]
\centering
\includegraphics[scale=1, angle=0]{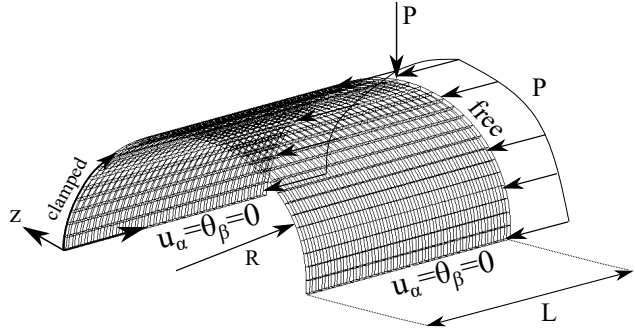}
\caption{A composite semi-cylindrical shell under compressive and transverse loads.}\label{fig 4702}
\end{figure}

The second analysis example concerns a clamped composite semi-cylindrical shell under compressive and transverse loads. Regarding boundary conditions, the vertical deflection and the rotation about the $\beta$-axis are restrained along its longitudinal edges.
This laminated composite cylindrical shell considering the stacking sequence [$90^\circ$, $0^\circ$, $90^\circ$] is reported in Fig. \ref{fig 4702}. In the same figure, the boundary conditions and applied loads are shown.
The investigated model has the following geometric data:
$L$ = 3.048 m, $R_{\alpha}$ = 1.016 m, and the thickness equal to 0.03 m. Each layer is made up of an orthotropic material with the following properties: $E_{L}$ = 2068.5$\times$10$^4$ N/m$^2$, $E_{T}$ = 517.125$\times$10$^4$ N/m$^2$, $G_{LT}$ = 795.6$\times$10$^4$ N/m$^2$ and Poisson's ratio $\nu_{LT}$ = $\nu_{TT}$ = 0.3.

\begin{figure}[htbp]
\centering
\includegraphics[scale=0.45, angle=0]{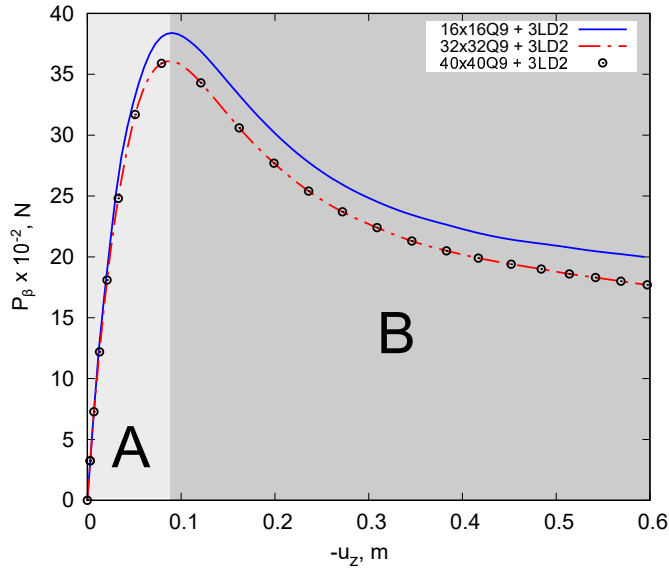}
\caption{Convergence study of the nonlinear response curves for the composite semi-cylindrical shell under compressive and transverse loads. Comparison of different in-plane mesh approximations. P$_{0}$/P$_{{\beta}_{0}}$ = 0.1515} \label{fig 060}
\end{figure}

First, convergence analyses of the in-plane mesh approximation are carried out to achieve accurate static nonlinear response evaluation. In the thickness direction, only one LD2 for each layer is adopted.
Figure \ref{fig 060} provides the transverse deflection at the loading point for various 2D CUF shell models, where the in-plane meshes from 256Q9 to 1600Q9 FEs are employed. The nonlinear response curves are split into two regions A and B. 
Moreover, the transverse displacement values for various models and loads, along with the DOFs, are reported in Table \ref{Table060}.
\begin{table}[htbp]
\centering
\begin{tabular}{cccccc}
\toprule
\toprule 
\multirow{3}{*}{CUF shell model}  & \multirow{3}{*}{DOFs}  &   \multicolumn{3}{c}{-$u_{z}$ [m]} \\
\cmidrule(l){3-5}
 & & \multicolumn{3}{c}{$[90^\circ, 0^\circ, 90^\circ]$} \\
\cmidrule(l){3-5}
 &  & 1800 N & 2500 N in A & 2500 N in B  \\
\midrule
16 x 16Q9 + 3LD2  & 22869  & 0.021 & 0.030 & 0.304  \\
32 x 32Q9 + 3LD2  & 88725  & 0.020 & 0.033 & 0.236  \\
40 x 40Q9 + 3LD2  & 137781 & 0.020 & 0.033 & 0.236  \\
\bottomrule
\bottomrule
\end{tabular} 
\caption{Equilibrium points of nonlinear response curves of the composite semi-cylindrical shell under compressive and transverse loads for various models and loads at the transverse loading point. P$_{0}$/P$_{\beta_{0}}$ = 0.1515}\label{Table060}
\end{table}
As evident from Fig. \ref{fig 060} and Table \ref{Table060}, the composite structure is well modelled through using 32$\times$32Q9 for the in-plane mesh approximation and only one LD2 for each layer in the thickness direction so as to achieve an accurate static nonlinear response analysis, which could be employed to show the various nonlinear strain terms effects.

\begin{figure}[htbp]
\centering
\includegraphics[scale=0.55, angle=0]{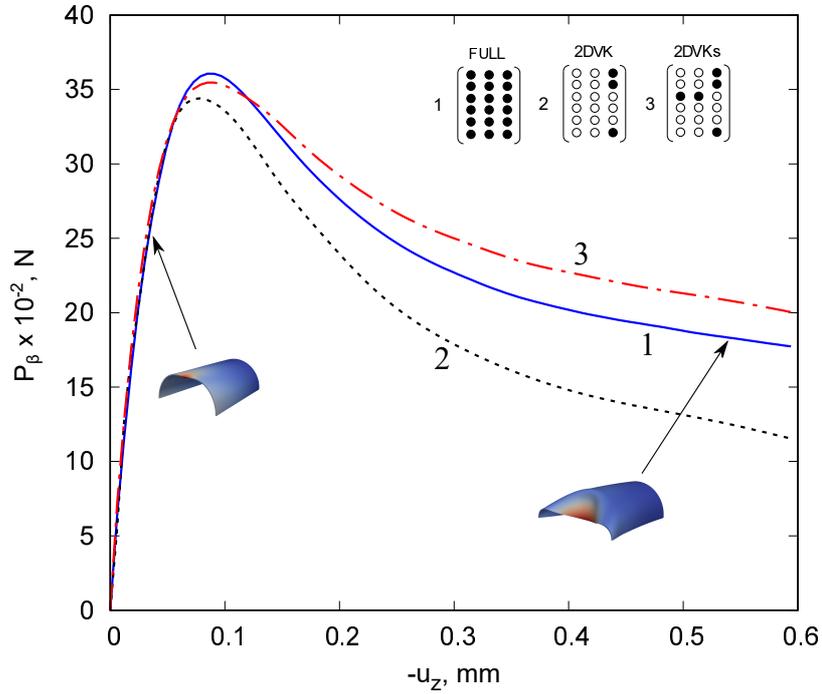}
\caption{Equilibrium curves evaluated at loading point of the composite semi-cylindrical shell under compressive and transverse loads for various geometrically nonlinear approximations. P$_{0}$/P$_{\beta_{0}}$ = 0.1515}\label{fig 4703}
\end{figure}

Figure \ref{fig 4703} shows the equilibrium curves of the composite semi-cylindrical shell under compressive and transverse loads based on different kinds of nonlinear strain approximations at the loading point.
In particular, the comparison between FULL, 2DVK and 2DVKs solutions is illustrated in the regime of moderate and large displacements/rotations.
It should be pointed out from Fig. \ref{fig 4703} that, compared with the FULL solution, the classical approximation of the von Kármán strains with nonlinear shear effects leads to less conservative results in the post-buckling regime. On the contrary, curve 2 (the 2DVK solution) provides a more conservative solution.
\subsection{Isotropic hinged cylindrical panel}
\begin{figure}[htbp]
\centering
\includegraphics[scale=0.3, angle=0]{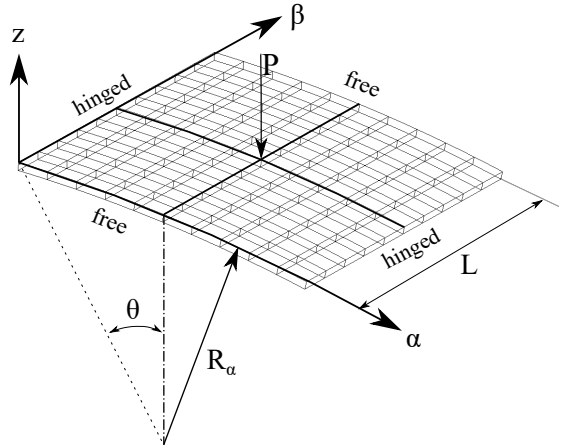}
\caption{A hinged cylindrical panel subjected to a central transverse load.}\label{fig 088}
\end{figure}

\begin{figure}[htbp]
\centering
\includegraphics[scale=0.45, angle=0]{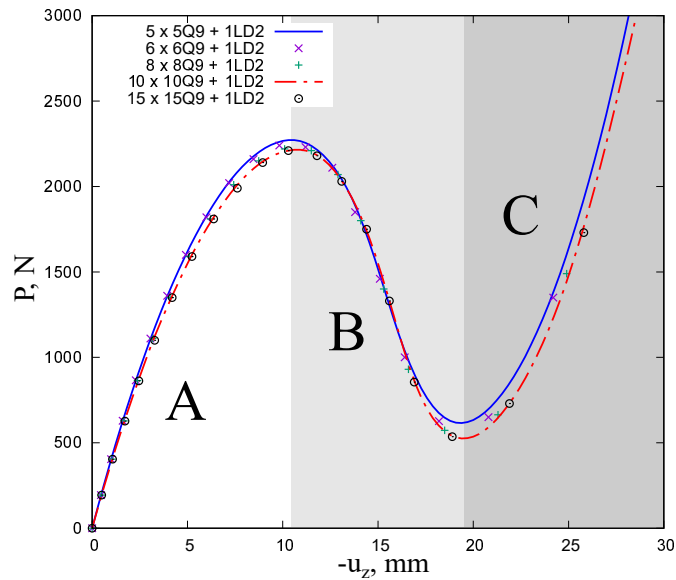}
\caption{Convergence study of the nonlinear response curves for the isotropic hinged cylindrical panel under a central transverse load. Comparison of different in-plane mesh approximations.}\label{fig 9991}
\end{figure}

A popular cylindrical panel undergoing snap-through under a transverse load is now considered.
As shown in Fig. \ref{fig 088}, the load condition is characterized by a central transverse force applied to the center of the structure.
The geometrical and material data are: $E$ = 3102.75 MPa and $\nu$ = 0.3, $L$ = 508 mm, $R_\alpha$ = 2540 mm, $\theta$ = 0.1 rad and thickness equal to 12.7 mm. Regarding boundary conditions, all nodal displacements are restrained along the hinged edges.
Before calculating the numerical results of different geometrically nonlinear models, we first perform a convergence study on the in-plane mesh approximation. In the thickness direction, only one LD2 is adopted.
Figure \ref{fig 9991} compares the transverse deflection for various 2D CUF shell model, where the in-plane mesh approximations from 25Q9 to 225 FEs are employed. The nonlinear response paths are split into three regions A, B and C. 
Furthermore, the transverse displacement values for various kinematic models and loads, along with the DOFs, are tabulated in Table \ref{Table32}.
\begin{table}[htbp]
\centering
\resizebox{\linewidth}{!}{
\begin{tabular}{ccccccccccc}
\toprule
\toprule 
\multirow{2}{*}{2D CUF model}  & \multirow{2}{*}{DOFs}  & \multicolumn{3}{c}{-u$_{z}$ [mm] in A} & \multicolumn{3}{c}{-u$_{z}$ [mm] in B} & \multicolumn{3}{c}{-u$_{z}$ [mm] in C}\\
\cmidrule(l){3-5} \cmidrule(l){6-8} \cmidrule(l){9-11}
 &  & 1000 N & 1500 N & 2000 N & 1000 N & 1500 N & 2000 N & 1000 N & 1500 N & 2000 N\\
\midrule
5 x 5Q9  + 1LD2  & 726 & 2.71 & 4.51 & 7.03 & 17.76 & 16.25 & 13.10 & 22.23 & 24.35 & 25.72 \\
6 x 6Q9  + 1LD2  & 1521 & 2.72 & 4.52  & 7.08 & 18.18 & 13.96 & 13.31 & 22.51 & 24.56 & 25.71 \\
8 x 8Q9  + 1LD2  & 2601 & 2.82 & 4.79  & 7.39 & 15.51 & 14.43 & 13.80 & 22.75 & 24.92 & 26.02 \\
10 x 10Q9 + 1LD2 & 3969 & 2.89 & 4.81  & 7.65 & 15.70 & 14.79 & 14.10 & 22.91 & 24.89 & 26.29 \\
15 x 15Q9 + 1LD2 & 8649 & 2.90 & 4.80 & 7.66 & 15.71 & 14.80 & 14.11 & 22.92 & 24.88 & 26.30\\
\bottomrule
\bottomrule
\end{tabular} 
}
\caption{Equilibrium points of nonlinear response curves of the isotropic hinged cylindrical panel for various in-plane mesh approximations and loads at the loading point.}\label{Table32}
\end{table}
As demonstrated from Fig. \ref{fig 9991} and Table \ref{Table32}, we could mode the hinged cylindrical panel by using 10$\times$10Q9 for the in-plane approximation to obtain an accurate nonlinear static response analysis, which will be utilized to show the effects of the different nonlinear strain terms effects.

\begin{figure}[htbp]
\centering
\includegraphics[scale=0.54, angle=0]{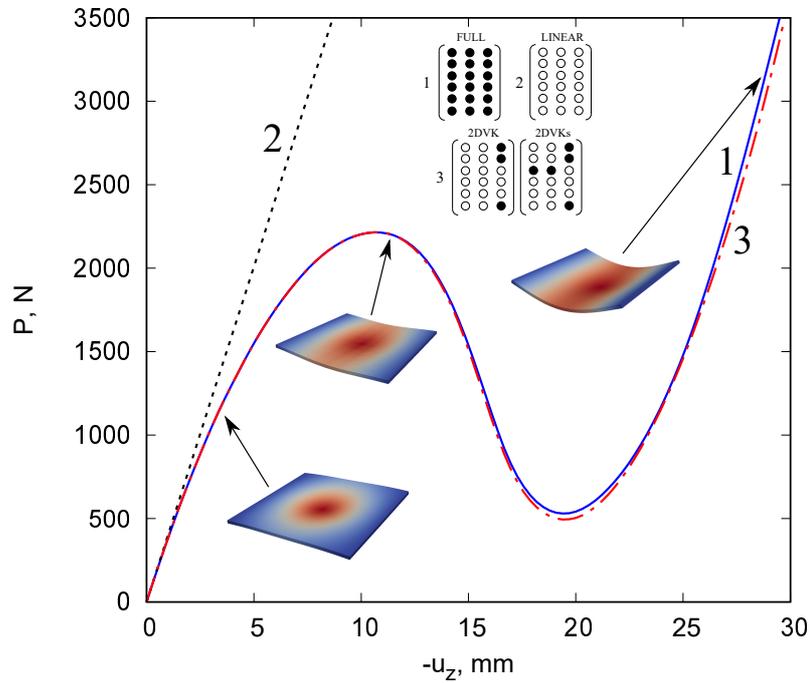}
\caption{Equilibrium curves evaluated at loading point of the isotropic hinged cylindrical panel for various geometrically nonlinear approximations.}\label{fig 033}
\end{figure}

Figure \ref{fig 033} shows the equilibrium curves of this popular hinged panel structure at the loading point. The linear, FULL nonlinear, 2DVK and 2DVKs solutions are illustrated in this figure.
For this case, the classical von Kármán nonlinear approximations (2DVK and 2DVKs) provide results similar to the 3D full nonlinear solution in the range of moderate displacements and of the same order of magnitude of the thickness.
\subsection{Isotropic and composite hinged cylindrical panel subjected to compressive and transverse loads}
\begin{figure}[htbp]
\centering
\includegraphics[scale=0.27, angle=0]{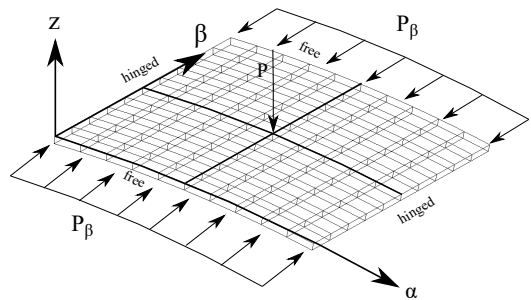}
\caption{A hinged cylindrical panel subjected to compressive and transverse loads.}\label{fig 7702}
\end{figure}

Now our interest is to compare the 3D full geometrically nonlinear CUF shell model and the classical von Kármán nonlinear theory for both isotropic and composite [$90^\circ$, $0^\circ$, $90^\circ$] hinged cylindrical panel under compressive and transverse loads.
The material data of isotropic panel are $E$ = 3102.75 MPa and $\nu$ = 0.3, whereas those of composite panel are: $E_{L}$ = 3330 MPa, $E_{T}$ = 1100 MPa, $G_{LT}$ = 660 MPa, $G_{TT}$ = 660 MPa, $\nu_{LT}$ = $\nu_{TT}$ = 0.25.
The geometrical characteristics, the boundary conditions and the loading conditions are illustrated in Fig. \ref{fig 7702}, with $L$ = 508 mm, $R_\alpha$ = 2540 mm, $\theta$ = 0.1 rad and thickness equal to 12.7 mm. In essence, all nodal displacements are restrained along the hinged edges. 

\begin{figure}[htbp]
\centering
\subfigure [Isotropic]
{\includegraphics[scale=0.4, angle=0]{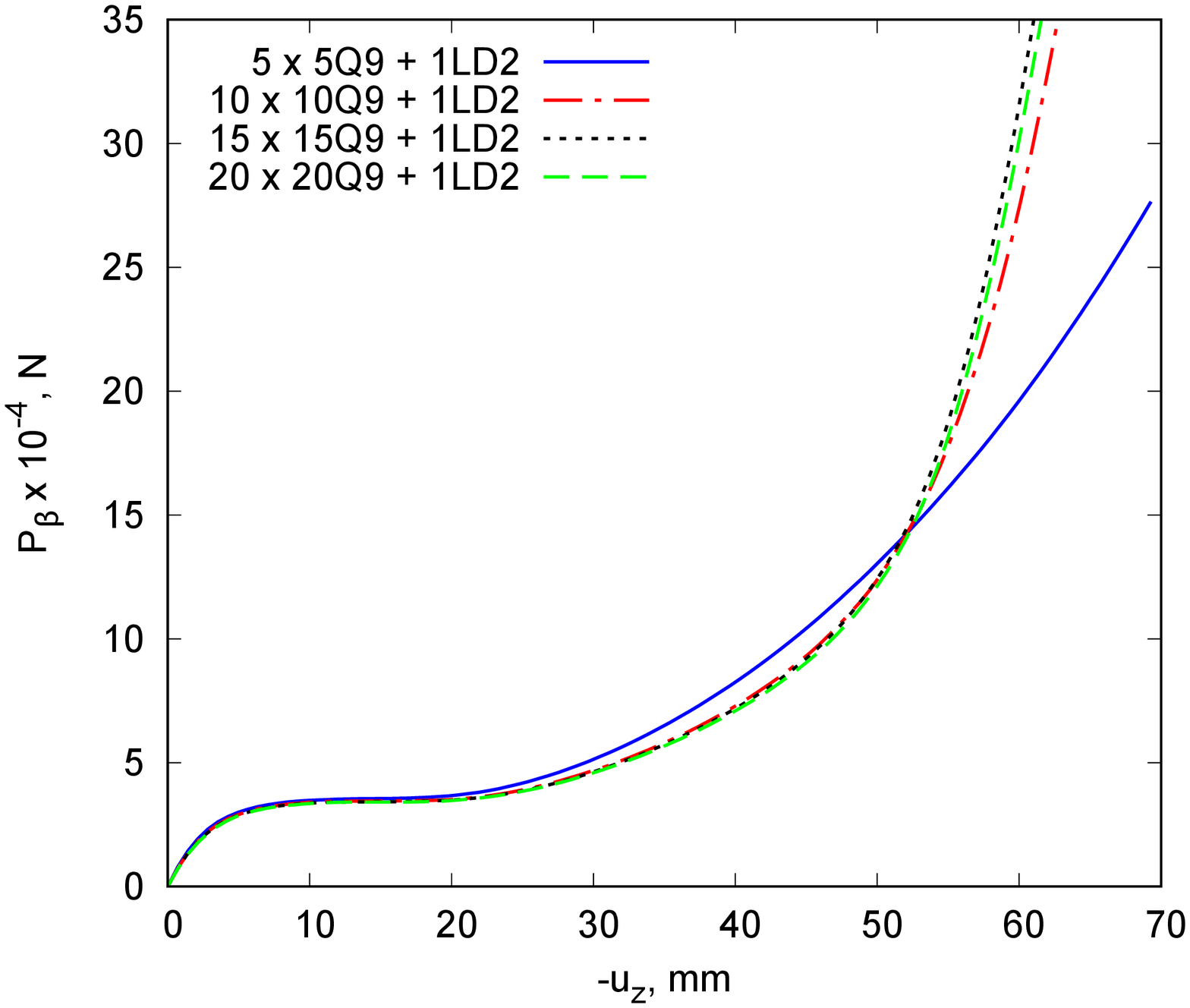}}
\subfigure [Composite]
{\includegraphics[scale=0.4, angle=0]{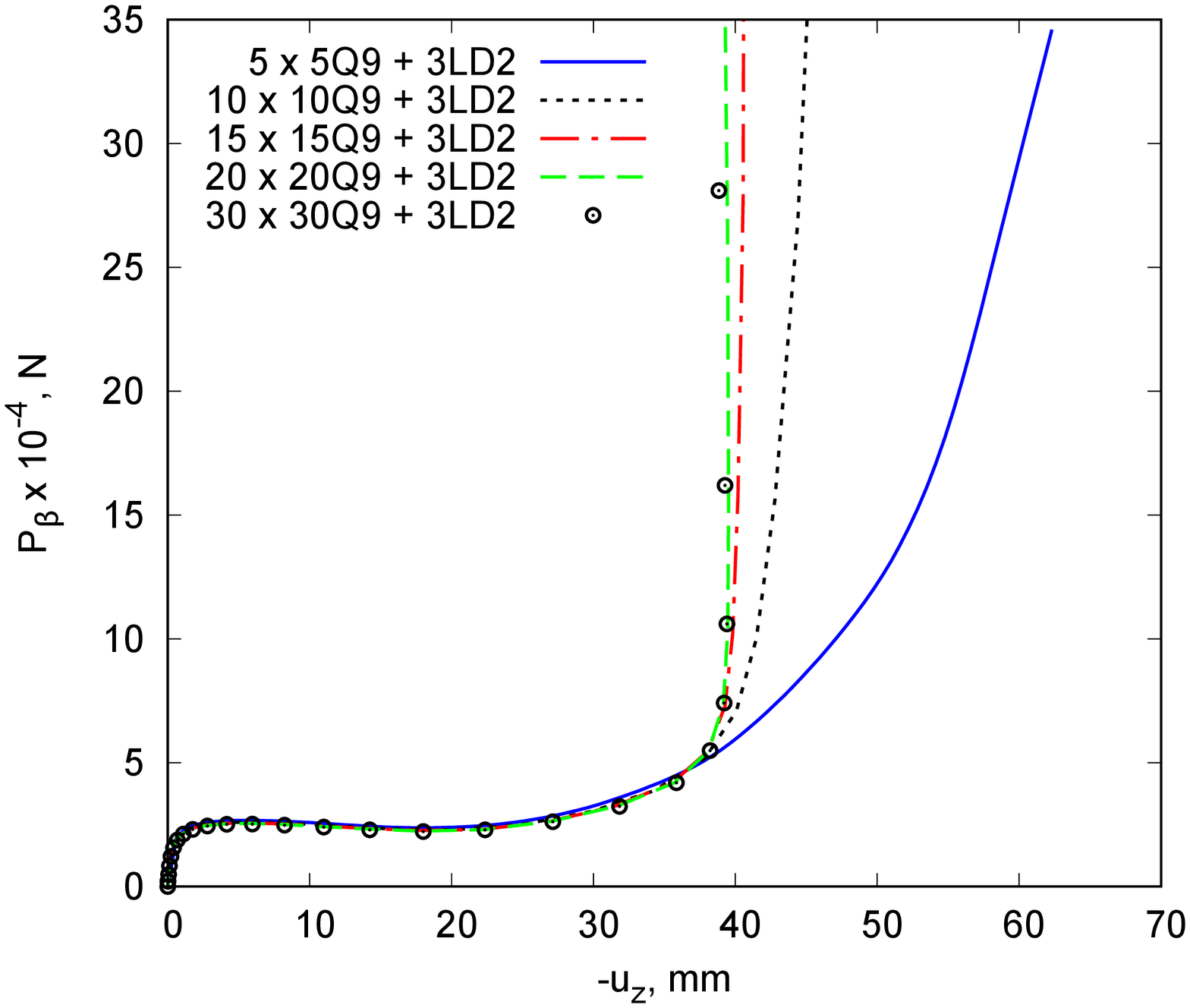}}
\caption{Convergence study of the nonlinear response curves for the isotropic and composite hinged cylindrical panel subjected to compressive and transverse loads. Comparison of different in-plane mesh approximations. P$_{0}$/P$_{\beta_{0}} = 0.0045$}\label{fig 504}
\end{figure}

First, convergence studies of the in-plane mesh approximation are performed to achieve an accurate static nonlinear response evaluation. In the thickness direction, only one LD2 per layer is adopted.
Figure \ref{fig 504} presents the transverse deflection versus P$_{\beta}$ for various 2D CUF shell models approximations considering P$_{0}$/$P_{\beta_{0}} = 0.0045$.  
Moreover, the transverse displacement values for various models and loads, along with the DOFs, are reported in Table \ref{Table66}.
\begin{table}[htbp]
\centering
\begin{tabular}{cccccccc}
\toprule
\toprule 
\multirow{1}{*}{CUF shell model}  & \multirow{2}{*}{DOFs}  & \multicolumn{2}{c}{-u$_{z}$ [mm]}\\
\cmidrule(l){1-1}
\cmidrule(l){3-4} 
$Isotropic$  &  & 10$\times$10$^4$ N & 30$\times$10$^4$ N \\
\midrule
5 x 5Q9  + 1LD2  & 1089 & 44.04 &  71.49 \\
10 x 10Q9 + 1LD2 & 3969 & 46.34 &  60.95  \\
15 x 15Q9 + 1LD2 & 8649 & 45.84 &  58.99   \\
20 x 20Q9 + 1LD2 & 15129 & 45.86 & 59.01    \\
\midrule
\midrule
\multirow{1}{*}{CUF shell model}  & \multirow{2}{*}{DOFs}  & \multicolumn{2}{c}{-u$_{z}$ [mm]}\\
\cmidrule(l){1-1}
\cmidrule(l){3-4} 
 $Composite$ &  & 10$\times$10$^4$ N & 30$\times$10$^4$ N \\
\midrule
5 x 5Q9   + 3LD2 & 2541  & 46.91 &  60.10  \\
10 x 10Q9 + 3LD2 & 9261  & 41.48 &  44.66   \\
15 x 15Q9 + 3LD2 & 20181 & 39.77 &  40.55   \\
20 x 20Q9 + 3LD2 & 35301 & 39.41 &  39.07  \\
30 x 30Q9 + 3LD2 & 78141 & 39.40 &  39.05   \\
\bottomrule
\bottomrule
\end{tabular} 
\caption{Equilibrium points of nonlinear response curves of the isotropic and composite hinged cylindrical panel under compressive and transverse loads for various in-plane mesh approximations and loads at the center of the structure. P$_{0}$/$P_{\beta_{0}} = 0.0045$}\label{Table66}
\end{table}
As seen from Fig. \ref{fig 504} and Table \ref{Table66}, the cylindrical panel under compressive and transverse loads is well modelled based on the in-plane mesh 15$\times$15Q9 for the isotropic model and 20$\times$20Q9 for the composite to conduct an accurate static response analysis.
 
\begin{figure}[htbp]
\centering
\subfigure [Isotropic]
{\includegraphics[scale=0.4, angle=0]{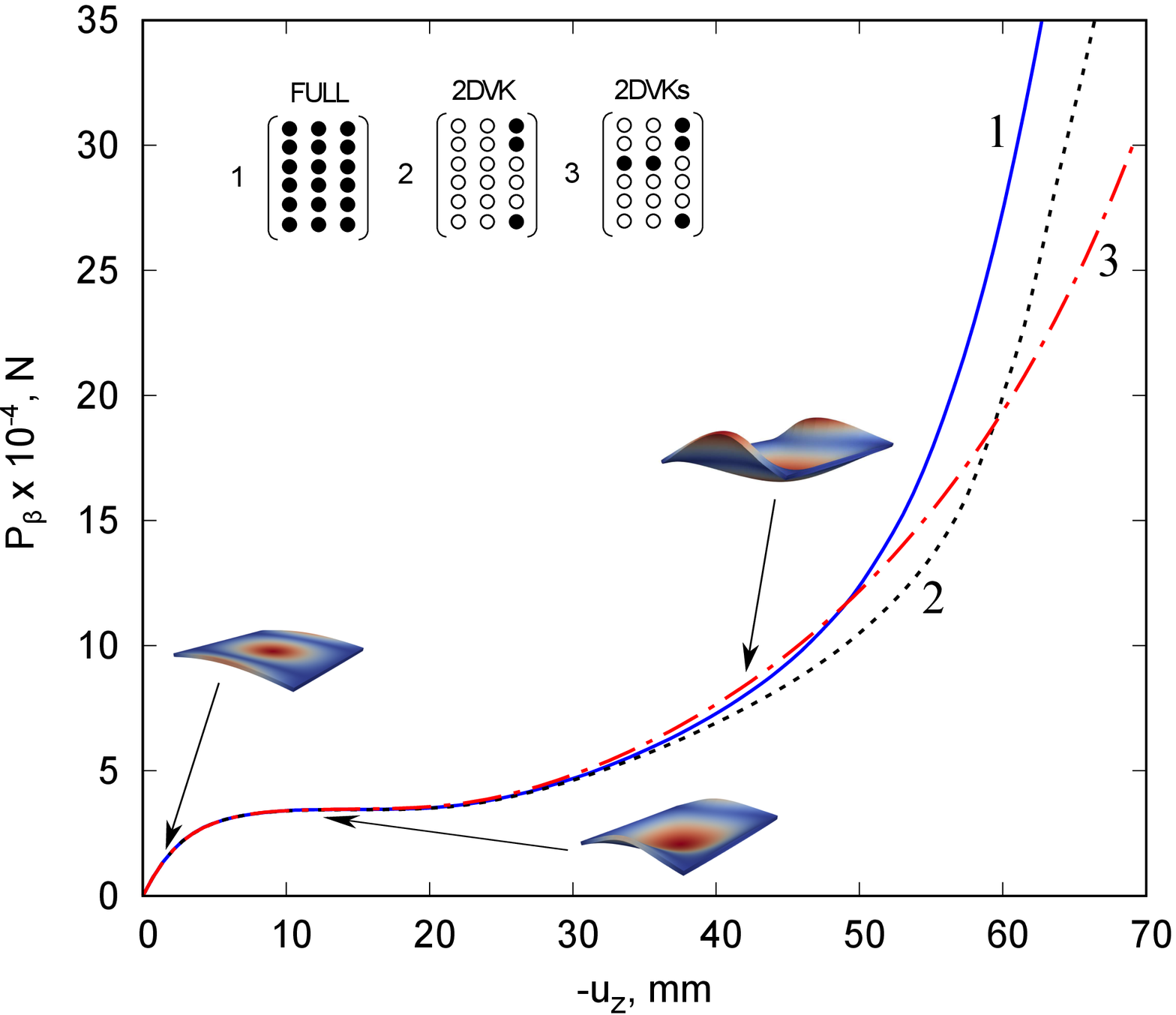}}
\subfigure [Composite]
{\includegraphics[scale=0.4, angle=0]{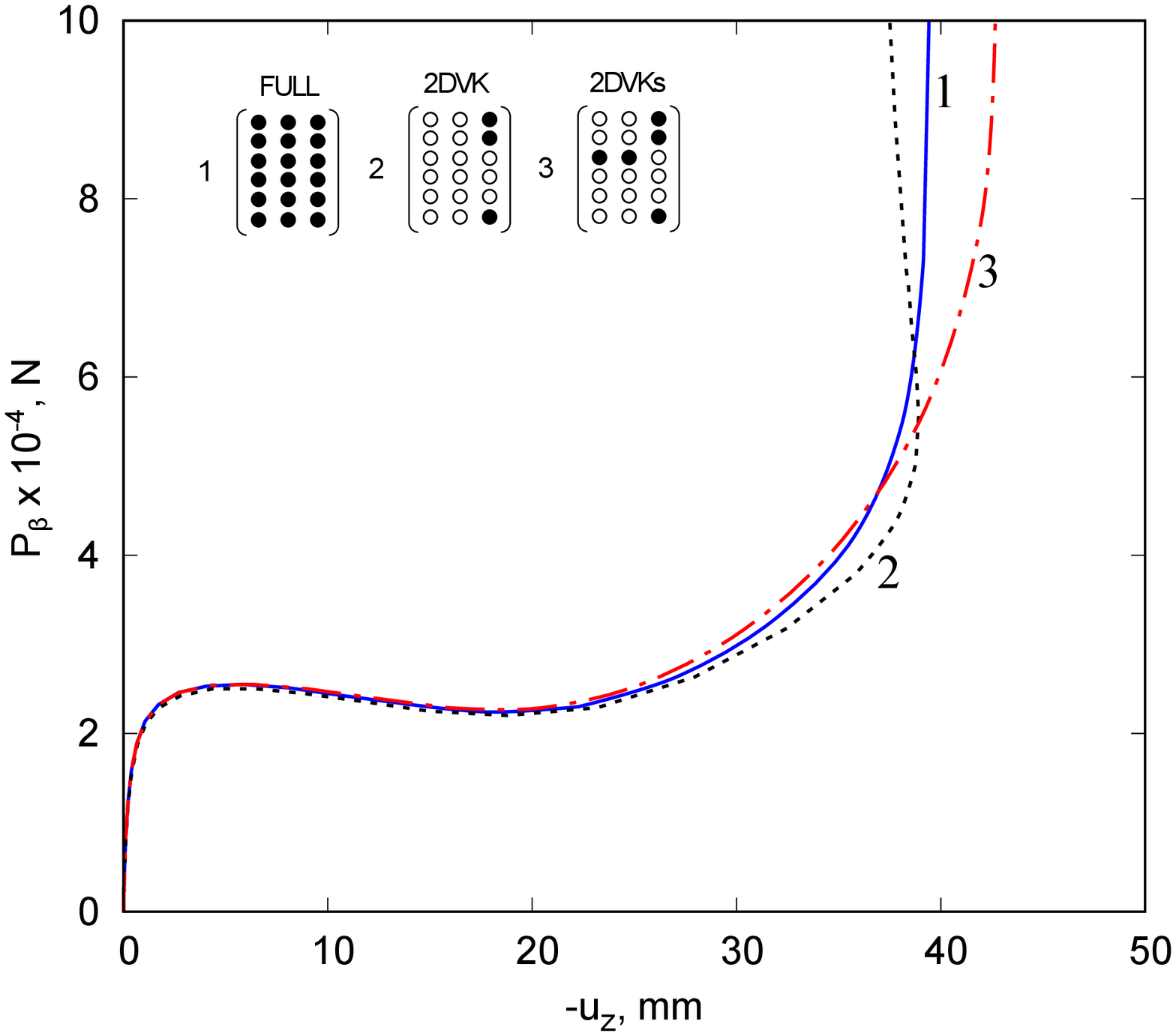}}
\caption{Equilibrium curves evaluated at the center of the (a) isotropic and (b) composite hinged cylindrical shell under compression and transverse loads for various geometrically nonlinear approximation. P$_{0}$/P$_{\beta_{0}} = 0.0045$}\label{fig 989}
\end{figure}

Figure \ref{fig 989} shows the equilibrium curves of the isotropic and composite hinged panel at the center of the structure. The 3D FULL nonlinear, 2DVK and 2DVKs solutions are illustrated in this figure. 
Obviously, for both isotropic and composite structures, the simplified 2D von Kármán models predict accurate results in the range of moderate displacements, while differences become more remarkable when large displacements are considered. Specifically, for the isotropic case the 2DVK is reliable up to about u$_z$ = 35 mm, whereas for the composite structure up to about u$_z$ = 25 mm.  The effective range of the 2DVKs predictions with nonlinear shear effects is larger than that of the 2DVK solutions for both isotropic and composite panels. Specifically, the 2DVKs prediction is effective up to about u$_z$ = 50 mm for isotropic case, while for composite case up to about u$_z$ = 35 mm.
%
%
%
\section{Conclusions} \label{V}
In the present article, nonlinear analyses of different popular shell structures in the deformed states of large displacements and rotations have been performed. 
These investigations have been conducted in the framework of the Carrera Unified Formulation (CUF). Thanks to its intrinsic scalable nature, the predictions  from three-dimensional (3D) full Green-Lagrange nonlinear strains to simplified nonlinear von Kármán strains are automatically obtained.
In fact, in the domain of CUF, the nonlinear governing equations and the related finite element (FE) arrays of any model are formulated through \textit{fundamental nuclei} (FNs), the structure of which are independent of the theory approximation order and the strain approximation considered.
We have adopted the Lagrange expansion (LE) to carry out detailed numerical evaluations for large displacements and rotations of isotropic and composite shell structures under transverse and compression loads. Various sets of nonlinear strain approximations have been analyzed, emphasizing the comparison between the 3D full Green-Lagrange nonlinear relations and the well-known two-dimensional (2D) von Kármán approximation.
In this context, the nonlinear equilibrium curves for each case have been illustrated and discussed.
As a conclusion to the proposed results, it is plausible to state that:
\begin{itemize}
\item This presented approach, based on the CUF, represent an efficient tool for comparing various geometrically nonlinear strain assumptions and kinematic approximation orders in an automatic way;
\item Simplifications of the 3D full Green-Lagrange (GL) nonlinear strains, such as the von Kármán theory, can provide good results for small/moderate displacements in the nonlinear regime, whereas in the regime of large displacements/rotations these simplified models present unacceptable results;
\item The 2D von Kármán strain approximation can be adopted for the analysis of pinched cylindrical shells and hinged cylindrical panels under transverse loads, but the 3D full geometrically nonlinear model should be employed for structures subjected to compressive loads.
The same considerations made for the snap-through instability case are also valid for the snap-back one, already studied in the works of Wu \textit{et al.} \cite{wu2019geometrically} and Shahmohammadi \textit{et al.} \cite{shahmohammadi2020geometrically}.
\end{itemize}
\section*{Declaration of competing interest}
Authors declare that they have no known competing financial interests or personal relationships that could appear to influence the work reported in this paper.
\section*{Author contributions}
\textbf{A. Pagani}: Supervision, Software, Writing - review $\&$ editing.
\textbf{R. Azzara}: Investigation, Visualization, Writing - original draft.
\textbf{E. Carrera}: Conceptualization, Funding acquisition, Methodology.
\textbf{B. Wu}: Visualization, Writing - review $\&$ editing.
\section*{Acknowledgments}
This work was supported by the European Union’s Horizon 2020 Research and Innovation Programme under the Marie Skłodowska-Curie Actions, Grant No. 896229 (B.W.).
%
%
%
\bibliographystyle{unsrt}
\bibliography{Shell_NL_terms}
%
%
\end{document}